\documentclass[twocolumn,eqsecnum,showpacs,twoside,pra]{revtex4}
\usepackage{mathrsfs}
\usepackage{amssymb, amsbsy, amsmath, latexsym, dsfont, array, layout, graphicx,mathrsfs,color}

\begin{document}

\title{Realization of a quantum walk in phase space using resonator-assisted double quantum dots}
\author{Zhihao Bian}
\affiliation{Department of Physics, Southeast University, Nanjing
211189, China}
\author{Hao Qin}
\affiliation{Department of Physics, Southeast University, Nanjing
211189, China}
\author{Xiang Zhan}
\affiliation{Department of Physics, Southeast University, Nanjing
211189, China}
\author{Rong Zhang}
\affiliation{Department of Physics, Southeast University, Nanjing
211189, China}
\author{Peng Xue}
\email{gnep.eux@gamil.com}
\affiliation{Department of Physics,
Southeast University, Nanjing 211189, China}
\affiliation{State Key
Laboratory of Precision Spectroscopy, East China Normal University,
Shanghai 200062, China}
\date{\today}

\begin{abstract}
We implement a quantum walk in phase space with a new mechanism
based on the superconducting resonator-assisted double quantum dots.
By analyzing the hybrid system, we obtain the necessary factors of
realization of a quantum walk in phase space: the walker, coin, coin
flipping and conditional phase shift. In order to implement the coin
flipping operator, we add a driving field to the resonator. The
interaction between the quantum dots and resonator field is used to
implement conditional phase shift. Furthermore, we show with
different driving fields the quantum walk in phase space exhibits a
ballistic behavior over 25 steps and numerically analyze the factors
which influence the spreading of the walker in phase space.

\end{abstract}


\maketitle

\section{Introduction}

Quantum walk (QW)~\cite{Aharonov, J.K, Luczak, Konno, Shikano,
Genske, Manouchehri,Rohde, Kitagawa, Matjeschk} is appealing as an
intuitively model in quantum computing and quantum
algorithms~\cite{Ambainis,Spielman,Whaley02,Kempe03}. Because it
exponentially speeds up the hitting time in glued tree graphs.
Furthermore, QW offers a quadratic gain over classical algorithms on
account of the diffusion spread (standard deviation), which is
proportional to elapsed time $t$, rather than $\sqrt{t}$ for the
classical random walk (RW)~\cite{Mackay,Brun}. Thus, the conception
of physical implementation of QW has become important and attract
more and more attention. Although realizations of QWs have been
proposed in different systems such as trapped
ions~\cite{Xue,Kirchmair,Schmitz}, photons~\cite{Bouwmeester, Do,
Zhang, Peruzzo, Perets, Sansoni, Jeong, Fedrizzi, Qin, Cassemiro,
Bao}, nuclear magnetic resonance~\cite{Chen,Kuznetsova} etc., the
solid state systems are attractive because of the stability and
expected scalability.

Recently, quantum computing with quantum dots has made a huge
progress~\cite{Culcer, Hari, Petta, Johnson, Solenov, Burkard, Lin, Golovach},
and the technique for coupling electrons associated with
semiconductor double-dot molecule to a microwave stripline resonator
has become more and more matured. Here we make use of the technology
and propose the implementation of a one-dimensional QW in phase
space with superconducting resonator-assisted quantum double-dot.
The walker is presented by a coplanar transmission line resonator
with a single mode and a two-level system--one electron shared by
double dots via tunneling serves as the quantum coin.

In our scheme the QW is executed with indirect flipping of the coin
via directly driving the resonator and allows controllable
decoherence over circles in phase space (PS) for observing the
transition between QW and RW~\cite{Burda,Lalumi,BC}. In next section
we give a brief introduction of the QW in PS. In Sec. III, we
implement QW via realizing the walker, coin, coin flipping and
conditional phase shift.  In addition to the numerical analysis
under the different driving fields, we we observe the ballistic
behavior of QW in PS and the QW-RW transmission with the influence
of decoherence introduced by the shift operation in the position
space.

\section{Brief introduction of QW in PS}  

Similar to the QW on a line in position space in which the walker
moves towards left or right based on the coin state, for QW on
circle in PS, the walker rotates either clockwise or
counter-clockwise along the circle in PS by the same amount, say an
angle $\Delta \theta$, with the choice of $\pm\Delta\theta$ strictly
random through the impulse, which is applied by a harmonic
oscillator.

In an ideal QW on a circle, the coin is replaced by a two-level
system with internal states $|0\rangle$ and $|1\rangle$. Here we
introduce the finite-dimensional orthogonal phase state
representation~\cite{Bartlett}
\begin{equation}
\label{eq:1}
|\theta_k=\frac{2\pi k}{d}\rangle=\frac{1}{\sqrt{d}}\sum^{d-1}_n {e^{in\theta_k}|n\rangle},k\in Z,
\end{equation}
where $|n\rangle$ is the Fock state. If the step size
$\Delta\theta=2\pi/d, d\in N$, then the walker always remains on the
circle with angular lattice spacing $\Delta\theta$. The walker walks
in PS with a state {$|\theta_k\rangle$} which can be decomposed into
the phase states. We introduce the rotation operator $\hat
R_m={e^{in\theta_m},m\in Z}$, then we have $\hat
R_m|\theta_k\rangle=|\theta_{k+m}\rangle$. We choose the Hadamard
operator
\begin{equation}
\label{eq:1}
H=\frac{1}{\sqrt{2}}{\left( {\begin{array}{*{20}c}
   1 & 1  \\
   1 & {-1}  \\
\end{array}} \right)}
\end{equation}
as the coin flipping operator and the unitary operation of one step
of the QW in PS is given by
\begin{equation}
\label{eq:1}
E=e^{in{\sigma_z}\Delta\theta}
\end{equation}
with the number operator on the walker state $n={a^\dag}a$, $a^\dag$
and $a$ are the creation and annihilation operators, respectively.
$\sigma_z={\left( {\begin{array}{*{20}c}
   1 & 0  \\
   0 & {-1}  \\
\end{array}} \right)}$ is one of the Pauli operators applied on the coin.

We define the initial state of walker+coin system as
\begin{equation}
\label{eq:1}
|\phi_0\rangle={|\phi_0\rangle}\otimes{|\varphi_0\rangle},
\end{equation}
where $|\varphi_0\rangle$ is the initial coin state. After $N$ steps, the system evolves to
\begin{equation}
\label{eq:1} |\phi_N\rangle=\left[E(I_\text{w}\otimes
H)\right]^N|\phi_0\rangle.
\end{equation}

The walker's phase distribution on a circle is
\begin{equation}
\label{eq:1} P(\theta)=\langle\theta|\rho_\text{w}|\theta\rangle
\end{equation}
with $d$ equally spaced values of $\theta=\theta_k$ and
$\rho_\text{w}=\text{Tr}_\text{c}|\phi_N\rangle\langle\phi_N|$ is
the reduced density matrix of the walker after tracing out the coin.

The standard deviation of the phase distribution satisfies,
$\sigma$, which is the symbol of the spreading of the QW, is linear
on time $t$. Therefore, in sufficiently short time, the relation
between phase spreading with time on a circle is a power law and
satisfies~\cite{Lalumi}
\begin{equation}
\label{eq:1}
\ln\sigma={\zeta}\ln t+\xi,
\end{equation}
with $\zeta=1$ for the QW and $\zeta=1/2$ for the RW.

\section{Physical implementation of QW}  

\subsection{Quantum coin and walker}  

Circuit Quantum Electrodynamics (QED) is a device which studies the
interaction between quantum particles and the quantized
electromagnetic modes inside a resonator. In this paper we consider
a hybrid QED system of superconducting resonator-assisted quantum
double-dot shown in Fig. 1a. One of the dots is capacitively coupled
to the resonator, and an electron shared by two adjacent dots
coupled via tunneling. The double dots can be modeled as a
double-well potential shown in Fig. 1b. Generally, we add an
external magnetic field along the axis $z$ to the double dots, which
the modest $B_z$ is $100$mT~\cite{Taylor}. Based on the external
magnetic field, there exists an energy difference between the two
potentials and the electron can tunnel between the two quantum dots.
We define the basis of qubits $|0\rangle$ for the electron appearing
in the left dot and $|1\rangle$ for the electron in the right dot.
From Fig. 1b we see the energy difference, namely energy cost which
moves the electron from the right dot to the left, between the
states $|0\rangle$ and $|1\rangle$ is $\Delta=T$, where $T$ is the
rate of electron tunneling in the different dots.

\begin{figure}
\includegraphics[width=8.5cm]{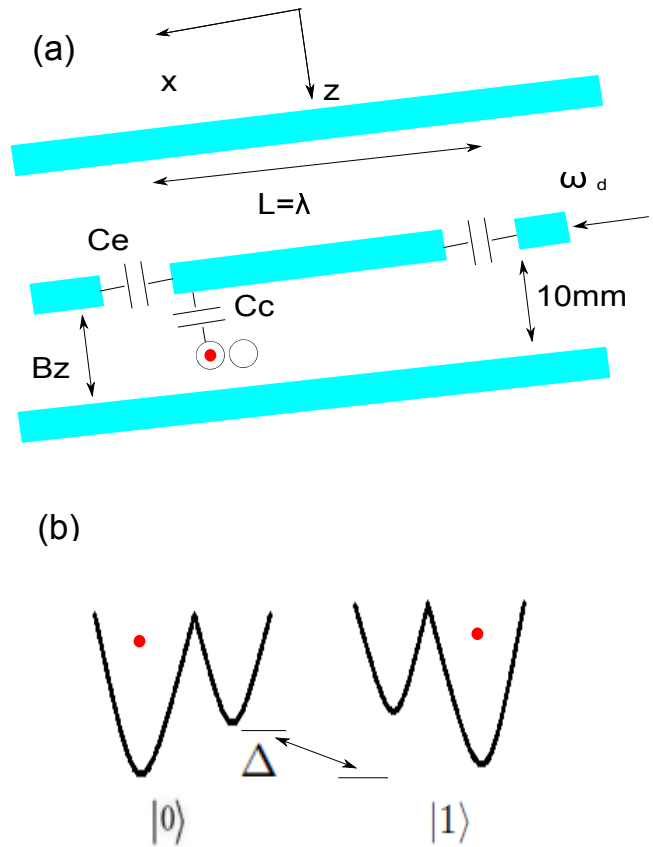}
\caption{(Color online.) (a) Experimental proposal for QW with
superconducting resonator-assisted quantum double-dot. The red dot
stands for the electron. The coupling between the resonator and the
double-dot can be switched on and off via the external electric
field along the $x$ axis. The superducting resonator is driven by a
field along the $x$ axis for implementation of coin flipping. (b)
The double-dot modeled as a double-well potential. The basis of the
qubit states represent the electron either in the left or right
potential via tunneling. The energy difference between the states
$|0\rangle$ and $|1\rangle$ is $\Delta$.}
 \label{figure:1}
\end{figure}

After adding the magnetic field to the quantum dots, the double-well
potential forms a circuit~\cite{Blais}. We just consider one
circumstance whether the electron locates in left dot or right, the
Hamiltonian describing the circuit is given by
\begin{equation}
\label{eq:1} \hat
H_\text{Q}=\sum_{N=0,1}E_c(N-N_g)^{2}|N\rangle\langle{N}|
-\Delta\left(|N+1\rangle\langle{N}|+H.C.\right),
\end{equation}
where $E_c=e^{2}/{2C_\text{tot}}$ is the charge energy,
$C_\text{tot}=C_g+C_J$ is the total capacitance in the circuit,
$C_J$ is the Josephson capacitance and $C_g$ is voltage biased from
a lead having capacitance to the circuit. $N_g={C_g}{V_g}/{2e}$ is
the gate charge which stands for the total polarization charge.
Restricting the gate charge to the range $N_g\in[0,1]$ by using the
voltage $V_g$, the Hamiltonian in Eq. (3.1) is rewritten as
\begin{equation}
\label{eq:1} \hat
H_\text{Q}=\frac{1}{2}E_c{(1-2N_g)}\sigma_z-\Delta{\sigma_x}.
\end{equation}

So far, we show the Hamiltonian of the double-well potential. From
Eq. (3.2) we see the double-well potential presenting the double-dot
system with effective electric fields along the $x$ and $z$
directions, and the system with internal states $|0\rangle$ and
$|1\rangle$ can be used as a two-level quantum coin.

Now we consider the circuit QED of double dots coupled to a
superconducting resonator. The dots are located in the center of the
resonator. If the oscillator mode of the resonator is coupled to the
double-dot, by using the coordinate system transformation $R=\left(
{\begin{array}{*{20}c}
   \cos\theta & {-\sin\theta}  \\
   \sin\theta & \cos\theta  \\
\end{array}} \right)$, and choosing $2N_g=1$, $\theta=\pi/2$,
the Hamiltonian of the interacting qubit and resonator system with the
rotating wave approximation takes the form ($\hbar=1$)
\begin{equation}
\label{eq:1} \hat
H_\text{JC}=\omega_ca^{\dagger}a+\frac{\Omega}{2}\sigma_z+g(a^{\dagger}\sigma_-+a\sigma_+),
\end{equation}
where $\omega_c$ is the frequency of the resonator,
$\Omega=\sqrt{{E_c}^{2}{(1-2N_g)}^{2}+4{\Delta}^{2}}$~\cite{Schoelkopf}
is the resonator induced energy splitting of the qubit,
$g=e{\frac{C_g}{C_\text{tot}}}\sqrt{\frac{\omega_c}{L}}$ is the
vacuum Rabi frequency. The form of Hamiltonian in Eq. (3.3) is the
well-known Jaynes-Cummings (JC) model~\cite{Jaynes}.

\subsection{Coin flipping and conditional phase shift operator}

In our hybrid system, the walker can be represented by the phase
state of the single mode of the resonator and the coin is the
two-level energy system. To implement the coin flipping operator, a
microwave time-dependent driving field is applied to the circuit QED
system with the form
\begin{equation}
\label{eq:1} \hat
H_\text{d}=\varepsilon(t)(a^{\dagger}e^{-i\omega_\text{d}t}+ae^{i\omega_\text{d}t}),
\end{equation}
where $\omega_\text{d}$ is the frequency of driving field. It is
easy to let $\varepsilon(t)$ to be a square wave, so $\varepsilon$
is a constant when the field is turned on, while it is zero when the
field is off. The Hamiltonian of the hybrid system, containing the
driving field, is
\begin{equation}
\label{eq:1} \hat H_\text{tot}=\hat H_\text{JC}+\hat H_\text{d}.
\end{equation}

In the dispersive regime,
\begin{equation}
\label{eq:1}
|\delta|=|\Omega-\omega_c|\gg{g},
\end{equation}
to calculate the effective Hamiltonian from Eq. (3.5), we introduce
the unitary transformation $\hat
S=\exp[g/\delta(a{\sigma_+}-a^\dag\sigma_-)]$, and use the
translated equation $ {\hat S^\dag}{\hat H_\text{tot}}{\hat
S}$~\cite{Blais,James}, the Hamiltonian in Eq. (3.5) turns into the
effective Hamiltonian of the whole system with driving field
\begin{equation}
\label{eq:1} \hat H_\text{eff}=\chi
a^{\dagger}a\sigma_z-\frac{1}{2}\delta_1\sigma_z-\delta_2a^{\dagger}a+\frac{1}{2}\Omega_2\sigma_x+\varepsilon(a^{\dagger}+a)
\end{equation}
with $\delta_1=\omega_d-\Omega$, $\delta_2=\omega_d-\omega_c$ , $\Omega_2=2g\varepsilon/\delta_2$, $\chi=g^2/\delta$.

The free evolution
\begin{equation}
\label{eq:1} \exp(-i\hat H_\text{eff} t)
\end{equation}
continues even when the driving field is off ($\varepsilon=0$).

The evolution of the hybrid system is described by the effective
Hamiltonian. The first term on the right-hand side in Eq. (3.7)
contains $a^{\dagger}a\sigma_z$, which proves an interrelated
relationship between the walker and coin, and makes the walker to
evolve along clockwise or counterclockwise at the same constant
angle with the orientation based on the state of the coin. The
second and third terms involve the operators $\sigma_z$ and
$a^{\dagger}a$, which represent the hence frequency of walker and
the energies for the coin, respectively. The fourth term, contains
$\sigma_x$, it can translate into the Hadamard coin flip by choosing
a suitable pulse time. The coefficient is proportional to the Rabi
frequency. The last term is the displacement in the position space
and pushes the walker off the circle in PS. Thus it also causes the
decoherence in the QW in PS.

\begin{figure}
\includegraphics[width=8.5cm]{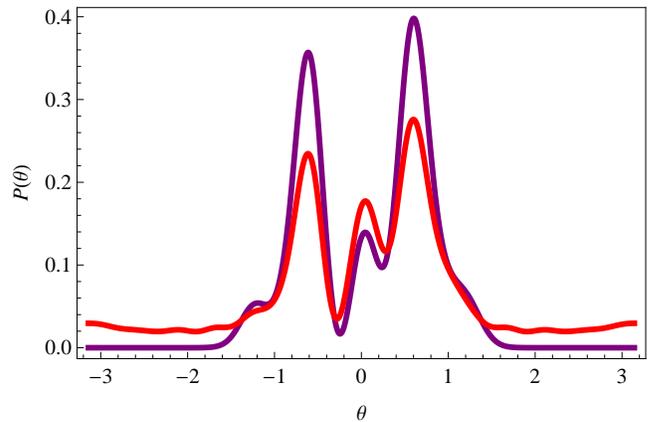}
\caption{(Color online.) The probability distribution of the walker
in PS at the 4th step with the initial walker state $|\alpha=
3\rangle$ and coin state $(|0\rangle+i|1\rangle)/\sqrt{2}$, the step
size $\Delta\theta=0.3$, and $\varepsilon=0.01$GHz in purple an
$\varepsilon=0.012$GHz in red respectively. The phase distribution
shows a ballistic behaviour of the QW in PS. With the strength of
the driving field increasing the decoherence introduced by
displacement in position space increases and the phase distribution
tends to be the combination of those of QW and RW.} \label{figure:2}
\end{figure}

\section{Numerical analysis}

Now all the factors that implementation of QW needs are fulfilled.
To make the the scheme working, the value of constant coefficient
$\varepsilon$ in last term in Eq. (3.7) which brings the
displacement in PS must be kept very small. Nevertheless, the Rabi
frequency $\Omega_2$ is proportional to the pulsed driving field
$\varepsilon$, so, however small $\varepsilon$ it is, we can choose
a suitable pulse time to translate the $\sigma_x$ into Hadamard coin
flip.

We choose the initial coin state as
\begin{equation}
\label{eq:1}
|\varphi_0\rangle=(|0\rangle+i|1\rangle)/\sqrt{2},
\end{equation}
the initial field in the resonator as coherent state
 \begin{equation}
\label{eq:1}
|\alpha\rangle=e^{-\frac{|\alpha|^2}{2}}\sum{\frac{\alpha^n}{\sqrt{n!}}|n\rangle},
\end{equation}
and access to the review, realistic system parameters are~\cite{Schuster,Gambetta}
\begin{equation}
\label{eq:1} (\Omega,\omega_c,g)/2\pi=(0.7,0.5,0.01)\text{GHz}.
\end{equation}

\begin{figure}
\includegraphics[width=8.cm]{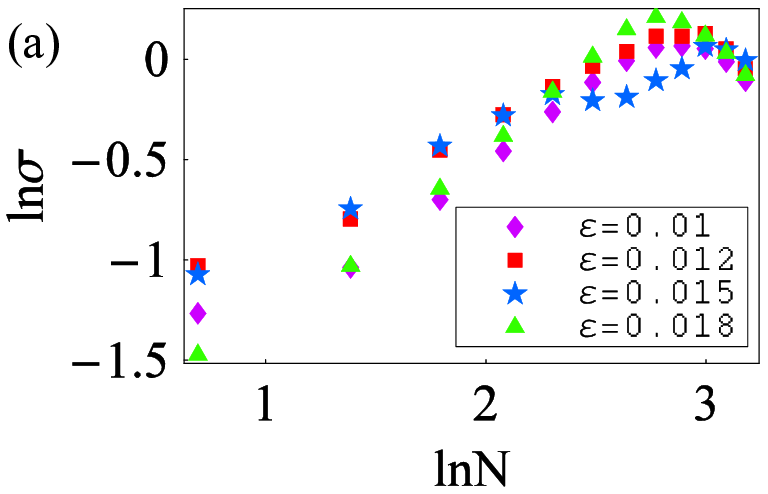}
\includegraphics[width=8.cm]{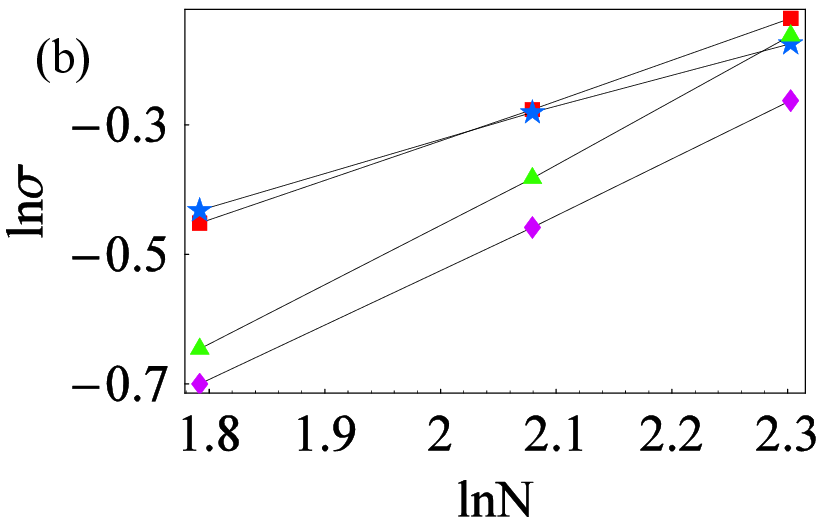}
\caption{(Color online.) (a) The ln-ln plot of the standard
deviation of the phase distribution $\sigma$  v.s. the step number
$N$ with the initial walker state $|\alpha=3\rangle$, coin state
$(|0\rangle+i|1\rangle)/\sqrt{2}$ and different strengths of the
field $\varepsilon=(0.01,0.012,0.015,0.018)$GHz, the step size in PS
$\Delta\theta=0.3$. The dots represent the standard deviation of
even steps up to $25$ steps. (b) The ln-ln plots with various
$\varepsilon$ show the slopes of the plots which represent the speed
of the walker spreading $\varsigma$.} \label{figure:3}
\end{figure}

By choosing the coherent state $|\alpha= 3\rangle$ and different
$\varepsilon$, we show the probability distribution of the walker in
PS at the 4th step with step size $\Delta\theta=0.3$ (Fig. 2). The
purple line represents the circumstances with $\varepsilon=0.01$GHz,
whereas, the red line stands for the circumstances with
$\varepsilon=0.012$GHz. From Fig. 2 we see the peak probabilities
with $\varepsilon=0.012$GHz is smaller than it with
$\varepsilon=0.01$GHz because the evolution of the last term in Eq.
(3.7) provides a displacement operator in position space and pushes
the walker off the circle in PS. Apart from this, the rate of
probability on the position $\theta=0.01$GHz in the three main peaks
with $\varepsilon=0.012$GHz is higher than it with
$\varepsilon=0.01$GHz, that is to say, the decoherence drives the
probability to the origin position, makes the distribution to
Gaussian distribution. These can be regarded as the decoherence on
the walker in PS and with decoherence increasing we can observe the
QW-RW transmission.

In Fig. 3a, we show the ln-ln plot of the standard deviation of the
probability distribution of the walker in PS with step size
$\Delta\theta=0.3$ only even steps without losing generation, the
purple, red, blue, green dots represent the standard deviation of
the phase distribution of QW in PS with different
$\varepsilon=(0.01,0.012,0.015,0.018)$GHz, respectively. With the
step number increasing, the wave function of the walker meets itself
on the circle in PS after $15$ steps. Thus after $15$ steps, the
curve of the ln-ln plot of standard deviation v.s. the step number
drops down. This shows difference from the QW on the line, though
for the first few steps QWs on both circle and line show the
ballistic behavior as expected. Overall, the slope of the points
becomes small with increasing of the constant $\varepsilon$. To
realize the discipline more intuitively, we cut out the standard
deviation after $6$, $8$, and $10$ steps and connect the points into
line of whole circumstances (Fig. 3b). The figure shows the slopes
of the line $\varsigma$ which is made up by the purple, red, blue,
green dots are about $1$, $0.89$, $0.64$, $0.53$ respectively. Using
Eq. (2.7) as a reference, the slop is $1$ for QW and $0.5$ for RW,
different slopes of the ln-ln plots show the different behavior of
the walker on circle in PS with different level of decoherence. The
decoherence is introduced by the driving field which leads the
walker to move in position space too. For QW in PS, it is equivalent
to decoherence. With the strength of the driving field increasing
the decoherence increases. Hence with decoherence increasing we see
the QW-RW transition. With the increasing of the value of
$\varepsilon$, the trend QW changes to RW is more obviously. So, in
order to keep the more prominent properties of QW, $\varepsilon$
must be small enough.

\section{Conclusion}

We show how a quantum walk in PS can be implemented in a quantum
quincunx created through superconducting resonator-assisted quantum
double dots and how interpolation from a quantum to a random walk is
implemented by controllable decoherence introduced by the
displacement of the walker in position space. Our scheme shows how a
QW with just one walker can be implemented in a realistic system.
The coin flipping operation is implemented by driving the resonator
directly, and at the same time the driving field also introduces the
displacement of the walker in position space and pushes the walker
off the circle in PS. Thus the displacement in position space is
equivalent to decoherence on the walker in PS which is controlled by
the strength of the driving field. With the strength of the driving
field increasing the decoherence increases and we observe the QW-RW
transition.

Although in our paper we make use of the decoherence introduced by
the driving field to show the transition from QW to RW, which is one
of the main points of our paper, for most of applications of QW it
requires quadratic enhancement of walker spreading. The decoherence
led by the driving field can be compensated with the method
in~\cite{Lalumi}. The displacement of the walker in position space
pushes the walker off the circle in PS by changing the mean photon
number of the resonator field. Hence we can adjust the pulse
duration each time according to the predicted mean photon number to
compensate the effect due to the displacement and obtain a perfect
QW in PS.

\begin{acknowledgments}
This work has been supported by the National Natural Science
Foundation of China under Grant No 11174052, the Open Fund from the
State Key Laboratory of Precision Spectroscopy of East China Normal
University. and the National Basic Research Development Program of
China (973 Program) under Grant No 2011CB921203.

\end{acknowledgments}


\begin{references}

\bibitem{Aharonov} Y. Aharonov, L. Davidovich, N. Zagury, Phys. Rev. A 48 (1993) 1687.
\bibitem{J.K} J.K. Asb\'{o}th, Phys. Rev. B 86 (2012) 195414 .
\bibitem{Luczak} A. W\'{o}jcik, T.  Luczak, P. Kurzy\'{n}ski, A. Grudka, T. Gdala, M. Bednarska-Bzdega, Phys. Rev. A 85 (2012) 012329.
\bibitem{Konno} N. Konno, Quantum Information Processing 9 (2010) 405.
\bibitem{Shikano} Y. Shikano, H. Katsura, Phys. Rev. E 82 (2010) 031122.
\bibitem{Genske} M. Genske, W. Alt, A. Steffen, A.H. Werner, R.F. Werner, D. Meschede, A. Alberti, Phys. Rev. Lett. 110 (2013) 190601.
\bibitem{Manouchehri} K. Manouchehri, J.B. Wang, Physical Implementation of Quantum
Walks, Springer-Verlag Berlin Heidelberg 2014
\bibitem{Rohde} P.P. Rohde, J.F. Fitzsimons, A. Gilchrist, Phys. Rev. Lett. 109 (2012) 150501.
\bibitem{Kitagawa} T. Kitagawa, M.S. Rudner, E. Berg, E. Demler, Phys. Rev. A 82 (2010) 033429.
\bibitem{Matjeschk} R. Matjeschk, A. Ahlbrecht, M. Enderlein, Ch. Cedzich, A.H. Werner, M. Keyl, T. Schaetz, R.F. Werner, Phys. Rev. Lett. 109 (2012) 240503.

\bibitem{Ambainis} A. Ambainis, International Journal of Quantum Information, 1 (2003) 507-518.
\bibitem{Spielman} A.M. Childs, R. Cleve, E. Deotto, E. Farhi, S. Gutmann, D.A. Spielman, Proc. 35th ACM Symposium on Theory of Computing(STOC 2003), pp. 59-68.
\bibitem{Whaley02} N. Shenvi, J. Kempe, K.B. Whaley, Phys. Rev. A  67 (2003) 052307.
\bibitem{Kempe03} J. Kempe, Contemporary Physics 44 (2003) 307-327.

\bibitem{Mackay} T.D. Mackay, S.D. Bartlett, L.T. Stephenson, B.C. Sanders, J. Phys. A 35 (2002) 2745-2753.
\bibitem{Brun} T.A. Brun, H.A. Carteret, A. Ambainis, Phys. Rev. Lett. 91 (2003) 130602.

\bibitem{Xue} P. Xue, B.C. Sanders, D. Leibfried, Phys. Rev. Lett. 103 (2009) 183602.
\bibitem{Kirchmair} F. Z\"{a}hringer, G. Kirchmair, R. Gerritsma, E. Solano, R. Blatt, C.F. Roos, Phys. Rev. Lett. 104 (2010) 100503.
\bibitem{Schmitz} H. Schmitz, R. Matjeschk, Ch. Schneider, J. Glueckert, M. Enderlein, T. Huber, T. Schaetz, Phys. Rev. Lett. 103 (2009) 090504.

\bibitem{Bouwmeester} D. Bouwmeester, I. Marzoli, G.P. Karman, W. Schleich, J.P. Woerdman, Phys. Rev. A 61 (1999) 013410.
\bibitem{Do} B. Do, M.L. Stohler, S. Balasubramanian, D.S. Elliott, C. Eash, E. Fischbach, M.A. Fischbach, A. Mills, B. Zwickl, J. Opt. Soc. Am. B 22 (2005) 499.
\bibitem{Zhang} P. Zhang, X.F. Ren, X.B. Zou, B.H. Liu, Y.F. Huang, G.C. Guo, Phys. Rev. A 75 (2007) 052310.
\bibitem{Peruzzo} A. Peruzzo, M. Lobino, J.C.F. Matthews, N. Matsuda, A. Politi, K. Poulios, X. Zhou, Y. Lahini, N. Ismail, K. W\"{o}rhoff, Y. Bromberg, Y. Silberberg, M.G. Thompson, J.L.O{'}Brien, Science 329 (2010) 1500.
\bibitem{Perets} H.B. Perets, Y. Lahini, F. Pozzi, M. Sorel, R. Morandotti, Y. Silberberg, Phys. Rev. Lett. 100 (2008) 170506.
\bibitem{Sansoni} L. Sansoni, F. Sciarrino, G. Vallone, P. Mataloni, A. Crespi, R. Ramponi, R. Osellame, Phys. Rev. Lett. 108 (2012) 010502.
\bibitem{Jeong} Y.C. Jeong, C. Di Franco, H.T. Lim, M.S. Kim, Y.H. Kim, Nature Communication 4 (2013) 2471.
\bibitem{Fedrizzi} M.A. Broome, A. Fedrizzi, B.P. Lanyon, I. Kassal, A. Aspuru-Guzik, A.G. White, Phys. Rev. Lett. 104 (2010) 153602.
\bibitem{Qin} P. Xue, H. Qin, B. Tang, B.C. Sanders, New J. Phys. 16 (2014) 053009.
\bibitem{Cassemiro} A. Schreiber, K.N. Cassemiro, V. Poto\v{c}ek, A.G\'{a}bris, P.J. Mosley, E. Andersson, I. Jex, C. Silberhorn,  Science 336 (2012) 55.
\bibitem{Bao} P. Xue, H. Qin, B. Tang, Scientific Reports, 4 (2014) 4825.

\bibitem{Chen} H.W. Chen, X. Kong, B. Chong, G. Qin, X.Y. Zhou, X.H. Peng, J.F. Du, Phys. Rev. A 83 (2011) 032314.
\bibitem{Kuznetsova} M.S. Kuznetsova, K. Flisinski, I. Ya. Gerlovin, M.Yu. Petrov, I.V. Ignatiev, S.Yu. Verbin, D.R. Yakovlev, D. Reuter, A.D. Wieck, M. Bayer, Phys. Rev. B 89 (2014) 125304.

\bibitem{Culcer} D. Culcer, A.L. Saraiva, B. Koiller, X.D. Hu, S. D. Sarma, Phys. Rev. Lett. 108(2012) 126804.
\bibitem{Hari} H.P. Paudel, M.N. Leuenberger, Phys. Rev. B 88 (2013) 085316.
\bibitem{Petta} J.R. Petta, A.C. Johnson, J.M. Taylor, E.A. Laird, A. Yacoby, M.D. Lukin, C.M. Marcus, M.P. Hanson, A.C. Gossard, Science 309 (2005) 2180.
\bibitem{Johnson} A.C. Johnson, Nature 435 (2005) 925.
\bibitem{Solenov} D. Solenov, S.E. Economou, T.L. Reinecke, Phys. Rev. B 87(2013) 035308.
\bibitem{Burkard}  G. Burkard, A. Imamoglu, Phys. Rev. B 74 (2006) 041307(R).
\bibitem{Lin}  Z.R. Lin, G.P. Guo, T. Tu, F.Y. Zhu, G.C. Guo, Phys. Rev. Lett. 101(2008) 230501.
\bibitem{Golovach} V.N. Golovach, M. Borhani, D. Loss, Phys. Rev. A 81 (2010) 022315.

\bibitem{Burda} Z. Burda, J. Duda, J.M. Luck, B. Waclaw, Phys. Rev. Lett. 102 (2009) 160602.
\bibitem{Lalumi} P. Xue, B.C. Sanders, A. Blais, K. Lalumi\'{e}re, Phys. Rev. A 78 (2008) 042334.
\bibitem{BC} P. Xue, B.C. Sanders, Phys. Rev. A 87 (2013) 022334.
\bibitem{Bartlett} B.C. Sanders, S.D. Bartlett, B. Tregenna, P.L. Knight, Phys. Rev. A 67 (2003) 042305.
\bibitem{Taylor} J.M. Taylor, H.A. Engel, W. D\"{ü}r, A. Yacoby, C.M. Marcus, P. Zoller, M.D. Lukin, Nature Physics 1 (2005) 177-183.
\bibitem{Blais} A. Blais, R.S. Huang, A. Wallra , S.M. Girvin, R.J. Schoelkopf, Phys. Rev. A 69 (2004) 062320.
\bibitem{Schoelkopf} R. Schoelkopf, A. Clerk, S. Girvin, K. Lehnert, M. Devoret, Kluwer Academic, Dordrecht (2003) Chap. 9, pp. 175-203.
\bibitem{Jaynes} E.T. Jaynes, F.W. Cummings, Proc. IEEE 51 (1963) 89.
\bibitem{James} D.F. James, J. Jerke, Canadian J. Phys. 85 (2007) 625.
\bibitem{Schuster} D.I. Schuster, A.A. Houck, J.A. Schreier, A. Wallraff, J.M.Gambetta, A. Blais, L.Frunzio, J. Majer, B. Johnson, M.H.Devoret, S.M. Girvin, R.J. Schoelkopf, Nature 445 (2007) 515.
\bibitem{Gambetta}  A. Blais, J. Gambetta, A. Wallraff, D.I. Schuster, S.M. Girvin, M.H. Devoret, R.J. Schoelkopf, Phys. Rev. A 75 (2007) 032329.

\end{references}
\end{document}